\documentclass[11pt,a4paper]{article}

\usepackage[T1]{fontenc}
\usepackage[utf8]{inputenc}
\usepackage{mathptmx}            
\usepackage[margin=2.5cm]{geometry}
\usepackage{amsmath,amssymb}
\usepackage{graphicx}
\usepackage{booktabs}
\usepackage{adjustbox}
\usepackage{microtype}
\usepackage[numbers,sort&compress]{natbib}
\usepackage[section]{placeins}
\usepackage{caption}
\usepackage{subcaption}
\usepackage{authblk}
\usepackage[hidelinks]{hyperref}
\hypersetup{colorlinks=true,citecolor=blue,urlcolor=blue,linkcolor=blue}

\title{\textbf{Multi-Line Spectral Reconstruction via Optoelectronic Chromatic Dispersion with a Single Photodiode}}

\author[1,*]{Endalamaw Ewnu Kassa}
\author[1]{Ziv Glasser}
\author[1]{Uttama Kumar Saint}
\author[1]{Roi Yozevitch}
\author[1]{Shmuel Sternklar}
\affil[1]{Department of Electrical and Electronic Engineering, Ariel University, Ariel 40700, Israel}
\affil[*]{Corresponding author: \href{mailto:endalamaw.kassa@msmail.ariel.ac.il}{endalamaw.kassa@msmail.ariel.ac.il}}

\date{}

\begin{document}

\maketitle

\begin{abstract}
Multi-line spectral reconstruction with a single germanium photodiode is demonstrated using optoelectronic chromatic dispersion (OED), without dispersive optics, filters or detector arrays. The wavelength-dependent absorption depth produces carrier-transit delays that encode spectral information in the phase and amplitude of the radio-frequency (RF) photocurrent. The RF response at 15 modulation frequencies (0.1--1.5~MHz), combined with the scalar DC photovoltage, forms a 31-element feature vector that machine learning maps to the spectrum. Under nested, wavelength-grouped cross-validation across the C and L bands (1530--1610~nm) and seven optical-power levels, Gaussian process regression reconstructs the wavelength of a single spectral line with a root-mean-square error (RMSE) of 0.38~nm, compared with 0.65~nm for the best classical calibration; the RF phase carries most of the wavelength information, and the amplitude spectrum adds complementary information. For mixtures with randomly varied wavelengths and powers, the pooled RMSE of the estimated line positions is 2.0~nm for two-line spectra and 3.3~nm for three-line spectra, two to four times lower than a reference that ignores the measurement; an independent campaign confirms these results across sessions. To our knowledge, this is the first demonstration of multi-line spectral reconstruction with a single photodiode using OED information. A singular-value analysis of the measured response links the individual-line accuracy to the number of independent measurement directions, which identifies additional measurement channels as a clear route to higher resolution in compact, alignment-free and potentially chip-integrated spectroscopy.
\end{abstract}


\section{Introduction}
\begin{figure}[!htbp]
\centering
\includegraphics[width=0.8\textwidth]{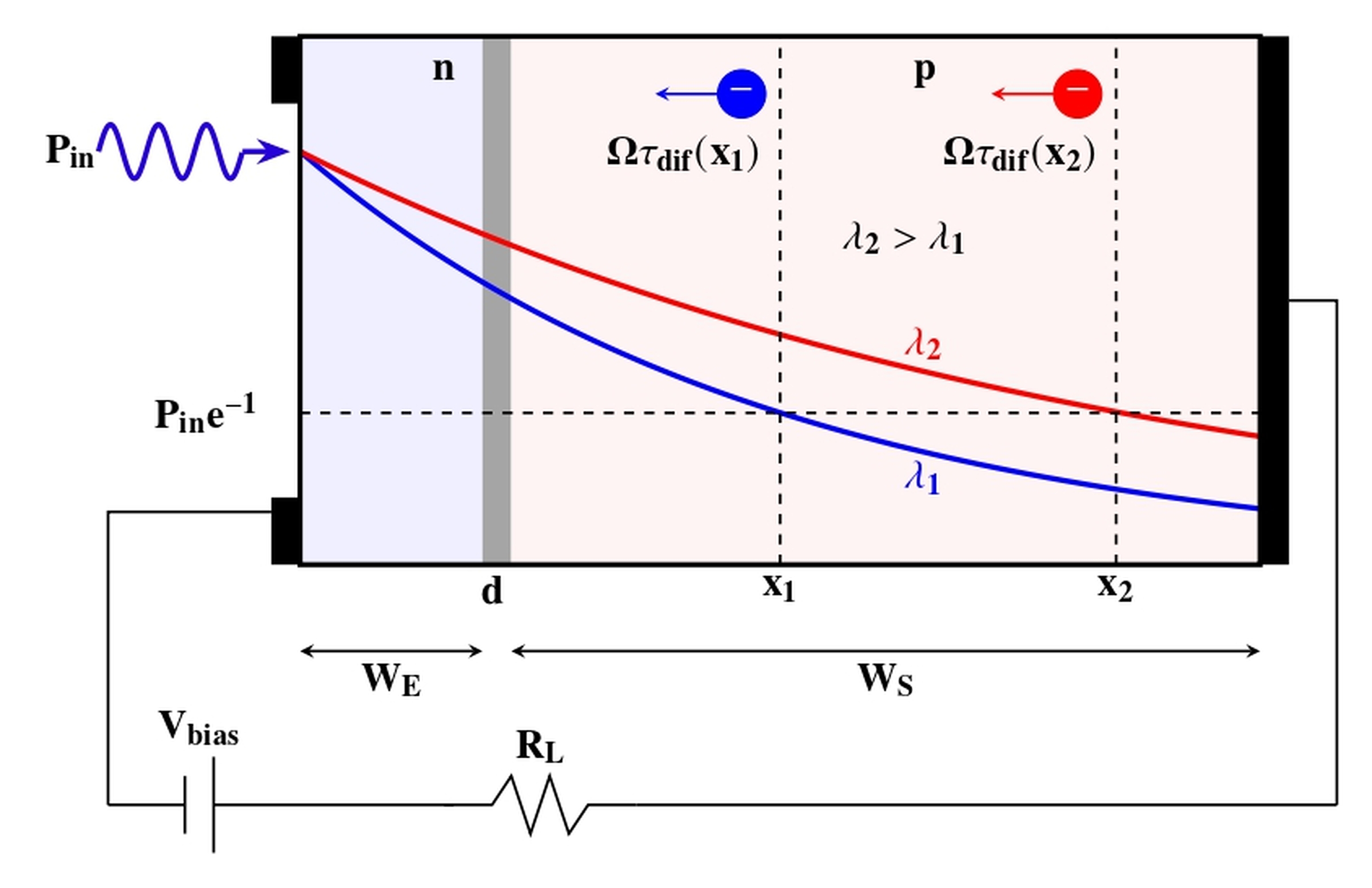}

\caption{Physical origin of optoelectronic chromatic dispersion in a
\textit{p--n} photodiode. Short-wavelength light ($\lambda_1$) is absorbed
near the surface, whereas longer-wavelength light ($\lambda_2$) penetrates
deeper into the substrate. The resulting differences in carrier diffusion
distance and transit time produce wavelength-dependent RF phase shifts in
the detected photocurrent.}
\label{fig:oed_schematic}

\end{figure}

Spectroscopy is a powerful tool for studying matter by providing detailed insights into its chemical composition, material properties, energy band structures, and environmental or biological processes \cite{prasad2024review,sinha2023spectroscopy,westermayr2025machine}. High-resolution spectral measurements are required for applications ranging from quantum material research and optical communications to medical diagnostics and remote sensing \cite{xue2024advances,yang2021miniaturization,dong2023recent}. Traditionally, spectrometers rely on gratings, prisms, or interferometer-based designs to separate light into spectral components \cite{guan2023review}. These systems can achieve high resolution and sensitivity but are often large, expensive, and complex \cite{li2022advances}. Their performance is also constrained by the optical path length of the instrument, making miniaturization challenging \cite{huang2024high,muir2024three,tang2024metasurface,zhang2025integrated}. To overcome these limitations, compact spectrometers have been developed using techniques such as multi-mode fibers \cite{amin2024multi}, disordered photonic chips \cite{redding2013compact}, planar photonic crystals \cite{wang2019single}, arrayed waveguide gratings \cite{gatkine2017arrayed}, micro-ring resonators \cite{zhang2022integrated}, and Mach-Zehnder interferometers \cite{zhang2025miniaturized,yang2021miniaturization}. These designs reduce the footprint of the device and improve the integration of the system; however, they often compromise the spectral accuracy, require precise optical alignment, or involve complex fabrication processes \cite{xue2024advances,yang2021miniaturization}. Computational or reconstructive spectrometers offer a new approach by moving the spectral discrimination from optics to algorithms. In these systems, wavelength-dependent information is encoded in measurable signals, such as spatial intensity patterns or multi-channel detector responses, and the input spectrum is reconstructed numerically or using machine learning (ML) models \cite{xue2024advances,su2025fast,zhang2022survey,zhang2025reconstructive,zhang2025integrated, ni2025machine}. Although this strategy allows smaller devices, most designs still require multiple detectors, precise optical alignment, or exhibit reduced optical throughput \cite{xue2024advances,li2022advances}. In this study, we present single-photodiode computational spectroscopy that utilizes optoelectronic chromatic dispersion (OED) as an intrinsic spectral encoder and reconstructs multi-line spectra. When a photodiode is illuminated with modulated light, it exhibits wavelength-dependent RF phase shifts in its photocurrent, which manifests as a type of dispersion known as an OED. Unlike the material dispersion \cite{saleh2019fundamentals,malitson1965interspecimen} found in all materials, the source of the OED is the wavelength-dependent penetration depth $\alpha(\lambda )$ and subsequent carrier migration, as shown in Fig.~\ref{fig:oed_schematic} \cite{glasser2021optoelectronic,dutta2024huge}. The photogenerated carriers migrate to the depletion region due to diffusion and drift, forming a photocurrent with a wavelength-dependent carrier transit time $\tau (\lambda )$, which results in an RF phase shift given by
\begin{equation}
\theta(\lambda) = 2\pi f \, \tau(\lambda).
\label{eq:phase_delay}
\end{equation}
For diffusion-dominant migration, the diffusion time is \(\tau_\mathrm{dif}(x_i) = L_{x_i}^2/D_i\), where \(L_{x_i}\) represents the distance to the depletion edge and \(D_i\) is the corresponding diffusion coefficient. A deeper penetration (e.g., \(\lambda_2\)) results in a larger \(\tau_\mathrm{dif}\) and phase shift. Under modulated illumination \(I_\text{in} = I_0(1 + m e^{i \Omega t})\), the AC photocurrent can be described as 

\begin{equation}
j_{\text{ac,tot}}(\lambda) = q_e\, m\, I_0\, e^{i \Omega t} \, F_{\text{tot}}(\lambda),
\end{equation}

where the complex transfer function can be expressed as \(F_{\text{tot}}(\lambda) = |F_{\text{tot}}(\lambda)| \, e^{-i \theta_{\text{tot}}(\lambda)}\)\cite{glasser2021optoelectronic,dutta2024utilization}. In this fashion, the spectrum is encoded based on the amplitude and phase of the modulated photocurrent. The OED sensitivity, defined as

\begin{equation}
S_\text{OED} = \frac{d\theta_\text{tot}}{d\lambda},
\end{equation}

is high in standard detectors, enabling high-resolution spectral sensing \cite{glasser2021optoelectronic,dutta2024large,mudgal2024large}.

\FloatBarrier
\FloatBarrier

By combining this simple physical encoding with machine learning models, the proposed method reconstructs optical spectra using only a single photodetector, without the need for dispersive optics, filter arrays, or moving parts. Machine-learning models learn and decode complex wavelength-dependent signals that are difficult to resolve with traditional designs, so the OED-ML computational spectrometer eases the conventional trade-off between size and performance and establishes OED-encoded carrier dynamics as a hardware-efficient foundation for portable, alignment-free spectroscopy. Using a Ge photodiode in the C and L bands (1530--1610~nm), we address five questions: how accurately the wavelength of a spectral line can be reconstructed under an evaluation that excludes spectral leakage; which part of the OED response carries the spectral information; what machine learning adds to classical calibration; how well the lines of a multi-line spectrum can be reconstructed; and what sets the resolution and operating range.

Under nested, wavelength-grouped cross-validation, Gaussian process regression (GPR) reconstructs the wavelength of a single spectral line with a root-mean-square error (RMSE) of 0.38~nm, against 0.65~nm for the best classical estimator (phase-only least-squares calibration) on the same folds; most of the gain comes from using the RF amplitude spectrum together with the phase. The detector's OED phase response carries substantially more wavelength-discriminating information than its amplitude response, because it does not depend on the optical power, and the amplitude shape is complementary. For two lines, the swept wavelength is reconstructed with 0.39~nm when the second line belongs to a set of calibrated wavelengths. For mixtures with randomly varied wavelengths and powers, the pooled RMSE of the estimated line positions is 2.0~nm for two-line spectra and 3.3~nm for three-line spectra, two to four times lower than a reference that ignores the measurement; an independent campaign with models trained in one session and tested in another confirms these results. To our knowledge, this is the first demonstration of multi-line spectral reconstruction with a single photodiode using OED information. A singular-value analysis of the measured response shows that the individual-line accuracy is governed by the number of independent response directions above the noise, so additional measurement channels, for example bias-tunable OED~\cite{dutta2024large} or detectors with different absorption edges~\cite{yoon2022miniaturized}, offer a clear route to higher spectral resolution.

Preliminary results from this work were presented in~\cite{kassa2026photodiode}. Here, we extend that study with a leakage-free evaluation, a feature ablation, a comparison with classical estimators, dual- and multi-line experiments, uncertainty quantification, and an experimental analysis of the resolution, spectral range, modulation and polarization dependence.

\section{Experimental Setup and Data Acquisition}
\label{sec:setup}

\begin{figure}[!htbp]
    \centering
    \includegraphics[width=0.8\linewidth]{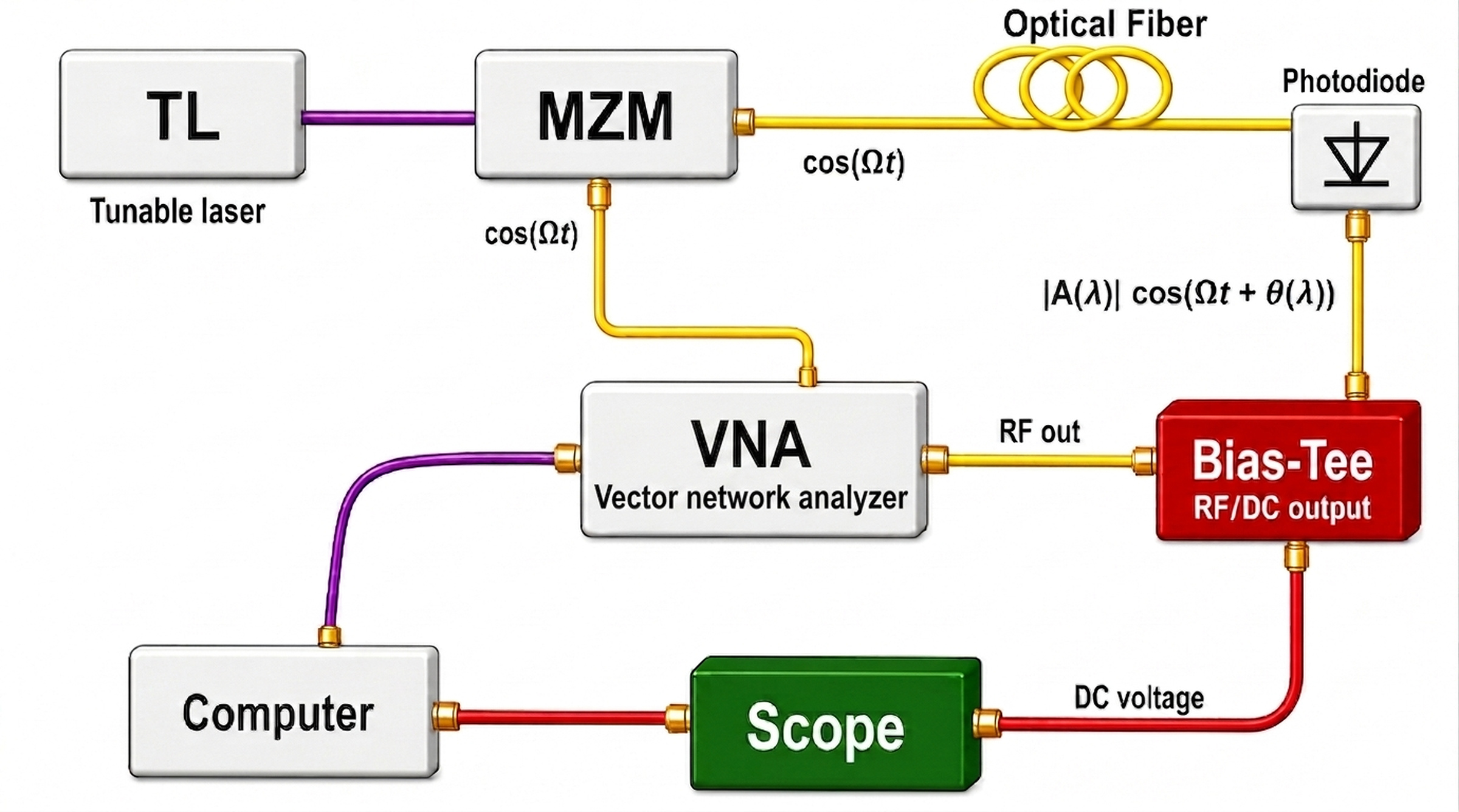}

    \caption{Experimental setup for OED measurement using the
    modulation phase-shift technique. A tunable laser is intensity-modulated
    by a Mach--Zehnder modulator and detected by the photodiode. A bias tee
    separates the RF signal, measured by the vector network analyzer, from
    the DC component, recorded by the oscilloscope. For two- and multi-wavelength spectra, several
    laser channels are combined in a coupler before the modulator. The instruments are
    synchronized and controlled by a computer.}
    \label{fig:setup}
\end{figure}

A schematic of the experimental setup is presented in Fig.~\ref{fig:setup}. A germanium 
PN photodiode (PDA50B-EC, Thorlabs) was illuminated by tunable laser sources (Keysight N7714A for the C band, Keysight N7711A for the L band; 1530--1610~nm). For a single-wavelength experiment, the laser output was coupled via polarization-maintaining (PM) fiber 
into a $\mathrm{LiNbO}_3$ Mach-Zehnder modulator (MZM) biased at its quadrature point 
($V_{\text{bias}} \approx 5~\text{V} \approx V_{\pi}/2$, where $V_\pi$ is the half-wave 
voltage of the modulator). This bias point was set at the start of each session to the middle of the measured modulator transfer curve (transmission 3~dB below its maximum), and the absence of a second harmonic in the detected RF signal was checked. The photodiode output was routed through a bias tee to isolate the AC and DC signal 
components. $S_{21}$ was measured with a Rohde \& Schwarz ZNLE4 vector network analyzer (VNA; 100~kHz--4.5~GHz): its source drove the MZM and its receiver, connected to the AC port, recorded the detector response while the RF frequency was swept from 0.1 to 1.5~MHz in 0.1-MHz steps (15 frequency points), with a 10-Hz IF bandwidth and a source power of $-10$~dBm. The RF modulation index at these low frequencies ($0.1$--$1.5\,\mathrm{MHz}$) is estimated as 
\(m=\pi V_{\mathrm{RF}}/(2V_{\pi})\ll 1\), confirming that the system operates well within the linear small-signal modulation regime. The DC 
port was simultaneously monitored by an oscilloscope (1~M$\Omega$ input impedance, 100~ms hardware averaging). The DC photovoltage, $V_{\mathrm{DC},i}$, is the time-averaged oscilloscope reading at the DC port of the bias tee; it is read 15 times at one operating point and the readings are averaged. These 15 readings are repetitions of one measurement, not values at the 15 modulation frequencies: their standard deviation is 0.7~mV (median), against a variation of 442~mV between samples (Supplementary Fig.~S10b). It is the output of the detector's transimpedance amplifier (PDA50B, 0~dB gain, 1.51~kV/A into a high-impedance load), read by the oscilloscope, and therefore proportional to the mean optical power on the detector; it is independent of the VNA modulation frequency, since both the mean optical power and the MZM quadrature bias point remain fixed during the VNA frequency sweep. 
Therefore, $V_{\mathrm{DC},i}$ is treated as a single scalar value for each measurement sample 
rather than a frequency-indexed quantity.
At each modulation frequency $f_i$ and wavelength $\lambda$, 
the system recorded the RF amplitude $A_i(\lambda)$, the measured phase $\theta_{\text{meas},i}(\lambda)$, 
and the DC voltage $V_{\text{DC}}$. To isolate the intrinsic device response, system phase calibration was performed at a 
reference wavelength $\lambda_{\text{ref}} = 1530$~nm. This reference accounts for 
frequency-dependent delays in the MZM driver, cables, and transimpedance circuitry. 
The measured phase is first unwrapped over frequency; subtracting the reference from all subsequent measurements of the same session gives the differential optoelectronic phase $
\theta_i(\lambda)=\theta_{\text{meas},i}(\lambda)-\theta_{\text{meas},i}(\lambda_{\text{ref}})$. At each wavelength, the system simultaneously sampled the response at 15~RF frequencies, 
forming a feature vector of 15~RF amplitudes and phases used for wavelength identification. 
The procedure was repeated at seven laser output power settings (4--10~mW, 1~mW steps; after the modulator and fiber losses the detector received about 0.05--1~mW, DC output 0.06--1.5~V) to assess feature stability against input power variation, giving 532 samples (76 wavelengths $\times$ 7 powers, one acquisition each, none missing or excluded). The wavelengths cover 1530--1565~nm (C-band laser) and 1571--1610~nm (L-band laser) in 1-nm steps; all set points, including 1609 and 1610~nm, were accepted by the lasers; 1566--1570~nm was not measured because it lies between the ranges of the two lasers. All measurements were conducted at room temperature.

The ground truth is the nominal set wavelength of the laser, and all errors in this work are relative to these set points. The datasheet specifies an absolute accuracy of $\pm$22~pm, a relative accuracy of $\pm$12~pm and a repeatability of $\pm$2.5~pm under the manufacturer's conditions (warm-up, ITU grid), which were not verified for every 1-nm set point. An optical spectrum analyzer measurement shows that the set point repeats to 0.5--0.6~pm after retuning, and the complete OED measurement repeats to 12--61~pm equivalent wavelength between independent sessions (Supplementary Section~S6). These uncertainties are compared with the estimation error in Section~\ref{sec:uncertainty}. For the dual-wavelength experiment, two independent laser channels were combined via a 
50:50~PM fiber coupler prior to modulation. The primary wavelength $\lambda_1$ was swept 
from 1530 to 1565~nm in 0.5-nm steps, while the secondary wavelength $\lambda_2$ was fixed at 1530, 1540, 
or 1550~nm (213 samples, one laser power setting of 4~mW, including three records with both channels at the same wavelength). Further campaigns, described in Section~\ref{sec:multiline}, combined two or three laser channels in the same way: a two-wavelength campaign with both wavelengths and their power ratio varied, a multi-line campaign with arbitrary spectra of one to three lines, and an independent multi-line campaign recorded in three sessions, in which a reference spectrum was measured periodically and the response of every acquisition was referred to the last preceding reference of the same session. In all cases the spectral lines are combined \emph{before} the modulator, so they share one fiber, one modulator and one detector, and their spatial overlap and the synchronization of their modulation are fixed by construction. A separate session (1540, 1550 and 1560~nm, 8~mW, 20 repeats each, with its own 1530-nm phase reference) was recorded to test transfer between sessions.

\FloatBarrier
\FloatBarrier

\section{Dataset and Machine Learning Algorithms}
\label{sec:sec_3.0}

\subsection{Dataset and Feature Representation}

\label{sec:sec_3.1}

Our dataset ${D}$ consists of $N$ experimentally acquired samples. For the single-wavelength experiments, the dataset comprises 532 samples (76 wavelengths $\times$ 7 power levels) collected across the 1530--1610~nm spectral range. For the dual-wavelength experiments, a dataset of 213 samples was acquired. Each sample pairs the photodiode's optoelectronic response with its ground-truth wavelength. At each wavelength $\lambda$ and optical power setting $P$, three physical observables were recorded; their names, symbols and dimensions are defined once in Table~\ref{tab:nomenclature} and used unchanged throughout the paper and the Supplementary Material: (i) the differential OED phase $\theta_i(\lambda, f_k)$ at each of the 15 modulation frequencies $f_k = 0.1k~\mathrm{MHz}$ $(k = 1,\ldots,15)$; (ii) the corresponding RF amplitude $A_i(\lambda, f_k)$, used as $\ln A$ so that the multiplicative dependence on optical power becomes an additive offset; and (iii) the DC photovoltage $V_{\mathrm{DC}}(\lambda, P)$, measured once per $(\lambda, P)$ step as the time-averaged oscilloscope reading at the bias-tee DC port, independent of the VNA modulation frequency (Section~\ref{sec:setup}). Formally, the feature vector for sample $i$ is defined as:

\begin{equation}
\mathbf{x}_i =
\left[
\theta_i(\lambda, f_1), \ldots, \theta_i(\lambda, f_{15}),\;
\ln A_i(\lambda, f_1), \ldots, \ln A_i(\lambda, f_{15}),\;
 \ln V_{\mathrm{DC}}(\lambda, P)
\right]^{\top}
\in \mathbb{R}^{31},
\label{eq:4}
\end{equation}
where $\theta_i(\lambda,f_k)$ and $A_i(\lambda,f_k)$ denote the RF total phase and amplitude at frequency $f_k$, and $V_{\mathrm{DC}}(\lambda, P)$ is the DC photovoltage measured once per $(\lambda, P)$ step as a single scalar quantity, independent of the modulation frequency. Like the amplitude, the DC voltage enters as its logarithm, so that the multiplicative dependence on optical power becomes an additive offset.
The dataset of $N$ samples is defined as
$
D = \left\{ (\mathbf{x}_i, y_i) \right\}_{i=1}^{N} \subset \mathbb{R}^{31} \times \mathbb{R}^{d},
$
where the target variable is the ground-truth wavelength: for the single-wavelength case $y_i = \lambda_i \in \mathbb{R}$ ($d=1$), and for the dual-wavelength case $y_i = (\lambda_{1,i}, \lambda_{2,i}) \in \mathbb{R}^{2}$ ($d=2$).

For clarity, we define two feature subsets that are frequently referenced throughout this work:
\begin{equation}
\mathbf{x}^{(\mathrm{OED})}_i =
\left[
\theta_i(\lambda, f_1), \ldots, \theta_i(\lambda, f_{15}),\;
\ln A_i(\lambda, f_1), \ldots, \ln A_i(\lambda, f_{15})
\right]^{\top}
\in \mathbb{R}^{30},
\end{equation}
\begin{equation}
\mathbf{x}^{(\mathrm{DC})}_i =
\left[
\ln V_{\mathrm{DC}}(\lambda, P)
\right]
\in \mathbb{R}.
\end{equation}
The complete 31-dimensional representation used for our full-model experiments is therefore
\begin{equation}
\mathbf{x}_i = \left[\,\mathbf{x}^{(\mathrm{OED})}_i \;\|\; \mathbf{x}^{(\mathrm{DC})}_i\,\right] \in \mathbb{R}^{31}.
\end{equation}
Feature subsets are named by combining the terms of Table~\ref{tab:nomenclature}, e.g.\ Phase + DC (16 features); the full set is OED + DC (31 features).

\begin{table}[!htbp]
\centering

\footnotesize
\begin{tabular}{p{3.0cm} p{9.4cm} c}
\toprule
Name (symbol) & Definition and measurement & Dimension\\
\midrule
Phase ($\theta$) & Differential RF phase, VNA $S_{21}$ at 0.1--1.5~MHz, unwrapped over frequency and referenced to the 1530-nm acquisition of the same session & 15\\
Amplitude ($\ln A$) & Natural logarithm of the RF amplitude $|S_{21}|$ at the same 15 frequencies & 15\\
DC ($\ln V_\mathrm{DC}$) & Natural logarithm of the DC photovoltage at the operating point of the sample; one scalar (mean of 15 repeated readings) & 1\\
\midrule
OED & Phase + Amplitude & 30\\
OED + DC & Phase + Amplitude + DC (full feature set) & 31\\
\bottomrule
\end{tabular}
\caption{Nomenclature of the OED features and feature sets.}
\label{tab:nomenclature}
\end{table}

To prevent spectral leakage, all samples measured at the same wavelength, at all seven power levels, are always assigned to the same fold, so a test wavelength is never seen at any power during training. Performance is estimated by nested, repeated, wavelength-grouped cross-validation. The outer loop has five folds, each withholding 15--16 of the 76 wavelengths, repeated with three random fold assignments (15 outer test folds; the test wavelengths of every fold are listed in Supplementary Table~S2). In the inner loop, hyper-parameters, and for GPR the kernel parameters by marginal likelihood, are selected by a grouped five-fold cross-validation inside each outer training set. Feature and target standardization, the classical calibration curves and all hyper-parameters are fitted on the outer training set only, and no feature selection is performed. The 1530-nm phase reference is a fixed per-frequency vector without fitted parameters that is subtracted from every sample of the session; within a session it is removed again by the feature standardization, so the cross-validation results do not depend on it (it matters only when data of different sessions are combined). Results are reported as the mean $\pm$ standard deviation of the RMSE over the 15 outer folds; the same folds are used for every model, classical estimator and feature subset. For the dual-wavelength data the folds are grouped by the swept $\lambda_1$, so every test record is an ordered channel pair $(\lambda_1,\lambda_2)$ absent from training; the reversed pair $(\lambda_2,\lambda_1)$ can occur in training when $\lambda_1\in\{1530,1540,1550\}$~nm. Stricter tests (withheld gaps and blocks of wavelengths, unseen power levels and an independent session) are reported in Section~\ref{sec:cv}.

\subsection{Machine Learning Models and Evaluation Metrics}
\label{sec:sec_3.2}

Four non-linear regressors are evaluated: Support Vector Regression (SVR), Random Forest (RF), Gradient Boosting Regression (GBR) and Gaussian Process Regression (GPR). GPR uses a Mat\'ern-5/2 kernel multiplied by a constant plus a white-noise term; the noise level is learned from the training data, and GPR also returns a predictive standard deviation, whose empirical coverage is reported in Section~\ref{sec:uncertainty}. Ridge Regression is kept only as a linear reference for this non-linear problem, labelled ``Ridge (linear ref.)'' in all tables and figures, together with second-order polynomial Ridge regression, and the gap between these models and GPR quantifies how non-linear the inverse problem is. For arbitrary multi-line spectra a deep neural network (DNN; an ensemble of multilayer perceptrons) is also evaluated (Section~\ref{sec:multiline}). The mathematical formulations and hyper-parameter search ranges are given in Supplementary Section~S1.

To judge what machine learning contributes, four classical (non-learned) estimators are evaluated with the same measurements and folds (details in Supplementary Section~S1): an analytical group-delay model, $\theta(\lambda,f)=2\pi f\,\tau(\lambda)$, inverted through the calibration curve $\tau(\lambda)$; a single-frequency calibration, i.e.\ the classical OED wavelength meter; a multi-frequency least-squares calibration, $\hat\lambda=\arg\min_\lambda\sum_{k}\big[\theta_\mathrm{meas}(f_k)-\theta_\mathrm{cal}(\lambda,f_k)\big]^2/\sigma_k^2$, which is the maximum-likelihood estimator for Gaussian phase noise; and a $k$-nearest-neighbour look-up table. For two-wavelength spectra, a physics-based inversion of the superposition of measured single-wavelength responses is also evaluated. The evaluation metrics, namely the root mean square error (RMSE), mean absolute error (MAE) and coefficient of determination ($R^2$), are defined in Supplementary Section~S2.

\section{Results and Discussion}

\subsection{Reconstruction Principle and Overall Design}

To reconstruct an unknown optical spectrum from the measured OED of a photodiode, we employed supervised machine-learning models that map the photodiode response directly to the spectral parameters. The experimental dataset used for training and evaluation is described in Section~\ref{sec:sec_3.1}. The reconstruction uses the wavelength-dependent response of the photodiode. Incident light generates a DC voltage and an RF signal whose amplitude and phase (\(\theta\)) vary with the wavelength, forming a spectral fingerprint. Under RF-modulated illumination, photons of different wavelengths are absorbed at different depths within the photodiode, resulting in wavelength-dependent carrier transit times and a characteristic phase shift (\(\theta\)) of the RF photocurrent (Fig.~\ref{fig:oed_schematic})~\cite{glasser2021optoelectronic}. These electrical signals intrinsically encode the optical wavelengths. Mathematically, for a single spectral line the reconstruction problem is formulated as a mapping from the measured photodiode response vector $\mathbf{x} \in \mathbb{R}^{31}$ to the optical wavelength $y \in \mathbb{R}$, where $y \equiv \lambda$. This relationship is expressed as $\hat\lambda = M(\mathbf{x})$, where $M(\cdot)$ denotes the machine-learning model and the input feature vector $\mathbf{x}$ is formulated as shown in Eq.~(\ref{eq:4}). Because the mapping $M(\mathbf{x})$ is non-linear and cannot be expressed in a closed analytical form, supervised regression models are trained using a dataset of $N_s$ experimentally acquired samples $\{(\mathbf{x}_i,\lambda_i)\}_{i=1}^{N_s}$. The model parameters are estimated by minimizing the empirical mean squared error between estimated and ground-truth wavelengths on the training folds. Each training pair \((\mathbf{x}_i, \lambda_i)\) consists of the measured photodiode response vector \(\mathbf{x}_i \in \mathbb{R}^{31}\) and its ground-truth optical wavelength \(\lambda_i\). The reconstructed wavelength \(\hat{\lambda}(\mathbf{x}_i)\) is obtained by evaluating the regression model \(M(\cdot)\), which learns the non-linear mapping between the OED features of the photodiode and the incident optical wavelength. For a spectrum of $N$ lines the same framework returns the ordered line wavelengths $(\hat\lambda_1,\ldots,\hat\lambda_N)$ and the photocurrent-weighted spectral centroid (Section~\ref{sec:multiline}). The learned models are compared with classical estimators that use the same data and folds (Section~\ref{sec:sec_3.2}).

\subsection{Statistical Characterization of the Feature Space}
\label{sec:features}

\begin{figure}[!htbp]
    \centering
    \includegraphics[width=\linewidth]{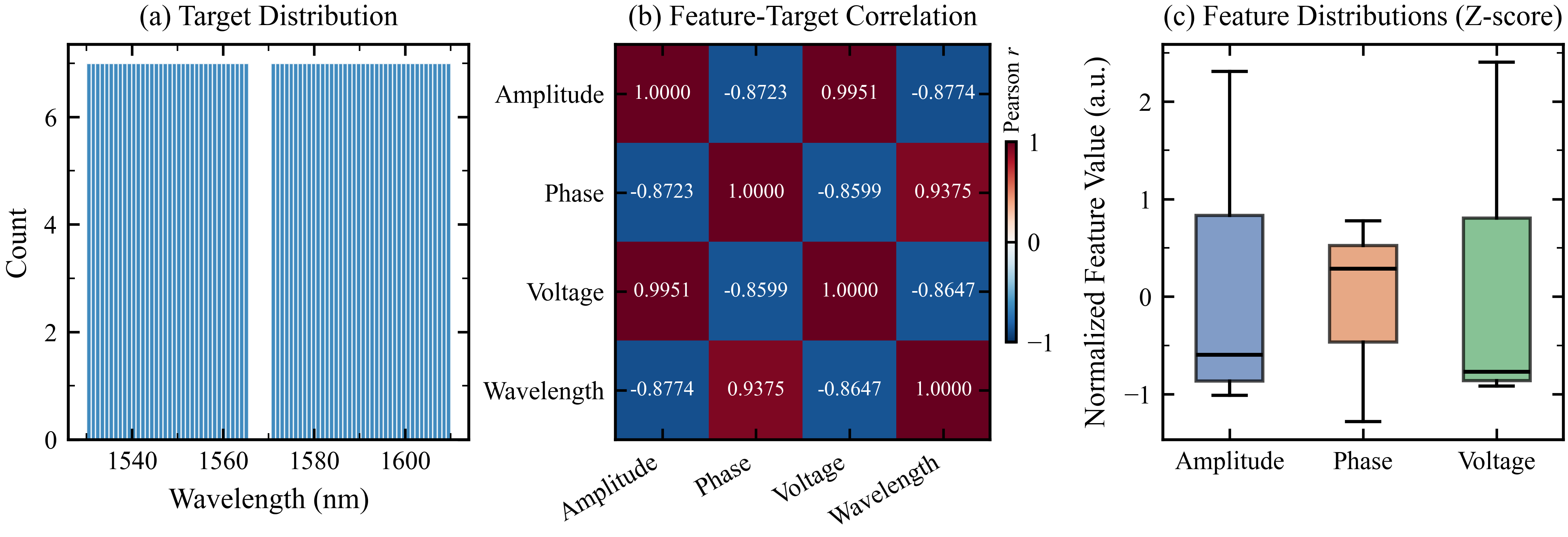}

    \caption{Dataset overview. (a) Distribution of target wavelengths across
    the C+L bands (76 wavelengths $\times$ 7 power levels = 532 samples). (b) Pearson correlations among the measured features and
    target wavelength. (c) Distributions of the amplitude, phase, and DC
    voltage features.}
    \label{fig:dataset_overview}
\end{figure}

The experimental dataset spans the C+L telecommunication bands ($1530$--$1610$~nm), and its statistical properties are summarized in Fig.~\ref{fig:dataset_overview}. As shown in Fig.~\ref{fig:dataset_overview}(a), the target wavelength distribution is uniform across the spectral range, with seven samples (one per power level) at each of the 76 wavelengths. The gap at 1566--1570~nm lies between the ranges of the C-band and L-band laser modules; the dataset remains balanced and free of significant spectral bias for machine-learning training and validation. The interdependence between the extracted features and the target wavelength was quantified using the Pearson correlation matrix of the frequency-averaged features shown in Fig.~\ref{fig:dataset_overview}(b). The RF amplitude and DC voltage features exhibit strong positive correlation ($r \approx 0.995$), reflecting their common dependence on the photogenerated carrier density and mean photocurrent of the photodiode. In contrast, the phase feature shows strong negative correlations with both amplitude ($r \approx -0.872$) and voltage ($r \approx -0.860$), indicating that the phase response evolves inversely with the intensity-related observables. The phase feature also shows the strongest positive correlation with wavelength ($r \approx 0.938$), confirming that wavelength-dependent dispersive phase shifts dominate the spectral encoding mechanism of the OED platform. All feature pairs exhibit high correlation magnitudes ($|r| \geq 0.86$), demonstrating substantial multicollinearity within the feature space. To improve numerical stability and ensure consistent convergence during model training, all features were standardized using Z-score normalization, fitted on the training folds only, as illustrated in Fig.~\ref{fig:dataset_overview}(c). The normalized distributions reveal that the amplitude and voltage features possess broader spreads with relatively lower median values, whereas the phase feature occupies a more symmetric and compact normalized range. This behaviour indicates that the phase response captures the dominant wavelength-dependent variance, while the amplitude and voltage primarily encode intensity-related information associated with the photocurrent generation process. The combination of a limited sample size and a strongly correlated feature space introduces a risk of overfitting. This risk is controlled by the nested, wavelength-grouped cross-validation of Section~\ref{sec:sec_3.1}, in which all preprocessing and tuning use training data only, and by the independent tests of Section~\ref{sec:cv}.

\FloatBarrier
\FloatBarrier

\subsection{Single-Wavelength Spectral Reconstruction}
\label{sec:single}

The central question that we address is whether the photodiode's modulated electrical output encodes sufficient chromatic information, through the OED, to replace a spectrometer and, if so, which features of that response carry the spectral signature most effectively. To address this, we compared three feature sets with five models under the nested, wavelength-grouped cross-validation (Table~\ref{tab:Table1}): (i) the DC voltage alone, (ii) the OED features, comprising the amplitude and phase at 15 modulation frequencies, and (iii) the full OED + DC set.

Models trained exclusively on the DC voltage reach at best $R^2 = 0.8345$ with an RMSE of 9.06~nm (GPR), an order of magnitude above the accuracy required for telecom spectroscopy. This result is physically intuitive: the DC photovoltage integrates carrier generation over the full absorption profile and, at seven power levels, mixes wavelength and power information, discarding the phase-resolved dynamics that distinguish one wavelength from another.

The dynamic OED features capture the wavelength-dependent modulation of carrier transport through the amplitude and phase of the photocurrent; unlike the DC voltage, they are intrinsically sensitive to the depth-dependent absorption profile, which shifts with wavelength. With the OED features, GPR reaches $0.371\pm0.163$~nm ($R^2 = 0.9997$) and SVR $0.554\pm0.273$~nm.

Adding the DC voltage gives $0.383\pm0.148$~nm for GPR, a difference of $+0.012$~nm that is not statistically significant (paired Wilcoxon signed-rank test over the same 15 outer folds, $p=0.93$). For a single line the OED response is therefore the spectral carrier, and the DC voltage is a redundant but harmless complement; its value appears for two lines at constant power (Section~\ref{sec:multiline}). Three further independent fold assignments give 0.375--0.399~nm, so the result does not depend on how the folds are drawn. A single 80/20 wavelength-grouped split can give a much lower value (0.179~nm, reached by 7 of 50 random splits), whereas the median over 50 splits (0.378~nm) agrees with the nested estimate (Supplementary Fig.~S7).

\FloatBarrier
\begin{table}[!htbp]
\centering
\renewcommand{\arraystretch}{1.25}
\setlength{\tabcolsep}{9pt}
\footnotesize
\begin{tabular}{l l c c c c}
\hline
\textbf{Feature Set} &
\textbf{Model} &
\textbf{Target} &
\textbf{$R^2$} &
\textbf{RMSE (nm)} &
\textbf{MAE (nm)} \\
\hline
\multicolumn{6}{c}{{DC Voltage}} \\
\hline
DC & SVR & $\lambda$ & 0.8270 & 9.219\,$\pm$\,0.624 & 7.603\\
DC & RF & $\lambda$ & 0.7694 & 10.682\,$\pm$\,0.552 & 8.519\\
DC & GBR & $\lambda$ & 0.7889 & 10.220\,$\pm$\,0.549 & 8.150\\
DC & GPR & $\lambda$ & 0.8345 & 9.057\,$\pm$\,0.511 & 7.490\\
DC & Ridge (linear ref.) & $\lambda$ & 0.8145 & 9.659\,$\pm$\,0.696 & 8.151\\
\hline
\multicolumn{6}{c}{{Phase \& Amplitude (OED)}} \\
\hline
OED & SVR & $\lambda$ & 0.9993 & 0.554\,$\pm$\,0.273 & 0.277\\
OED & RF & $\lambda$ & 0.9988 & 0.742\,$\pm$\,0.122 & 0.619\\
OED & GBR & $\lambda$ & 0.9985 & 0.830\,$\pm$\,0.118 & 0.716\\
OED & GPR & $\lambda$ & 0.9997 & 0.371\,$\pm$\,0.163 & 0.144\\
OED & Ridge (linear ref.) & $\lambda$ & 0.9866 & 2.549\,$\pm$\,0.469 & 1.923\\
\hline
\multicolumn{6}{c}{{OED + DC (full feature set)}} \\
\hline
OED + DC & SVR & $\lambda$ & 0.9994 & 0.518\,$\pm$\,0.217 & 0.266\\
OED + DC & RF & $\lambda$ & 0.9988 & 0.743\,$\pm$\,0.126 & 0.622\\
OED + DC & GBR & $\lambda$ & 0.9988 & 0.746\,$\pm$\,0.127 & 0.637\\
OED + DC & GPR & $\lambda$ & 0.9997 & 0.383\,$\pm$\,0.148 & 0.151\\
OED + DC & Ridge (linear ref.) & $\lambda$ & 0.9885 & 2.344\,$\pm$\,0.488 & 1.727\\
\hline
\end{tabular}
\caption{Single-wavelength reconstruction with three feature sets under nested, wavelength-grouped cross-validation (532 samples, 76 wavelengths, 4--10~mW). $R^2$, RMSE (mean $\pm$ s.d.) and MAE are averaged over the 15 outer folds; for RF and GBR the DC and OED rows use a reduced hyper-parameter grid. GPR with the OED features reaches sub-nanometer accuracy (about 0.38~nm); the OED + DC rows are identical to Table~\ref{tab:classical}.}
\label{tab:Table1}
\end{table}
\FloatBarrier

\begin{figure}[!htbp]
    \centering
    \adjustbox{max width=\linewidth}{\includegraphics{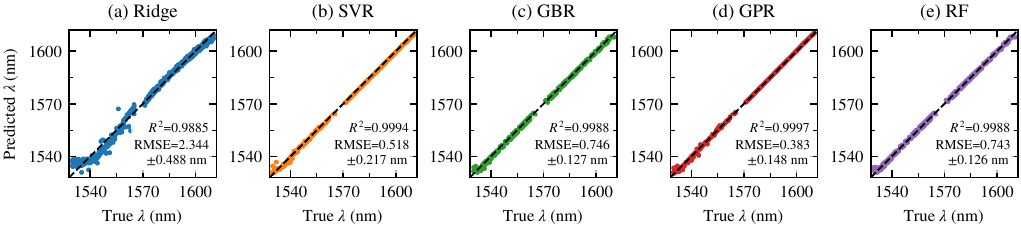}}
    \vspace{2mm}
    \adjustbox{max width=\linewidth}{\includegraphics{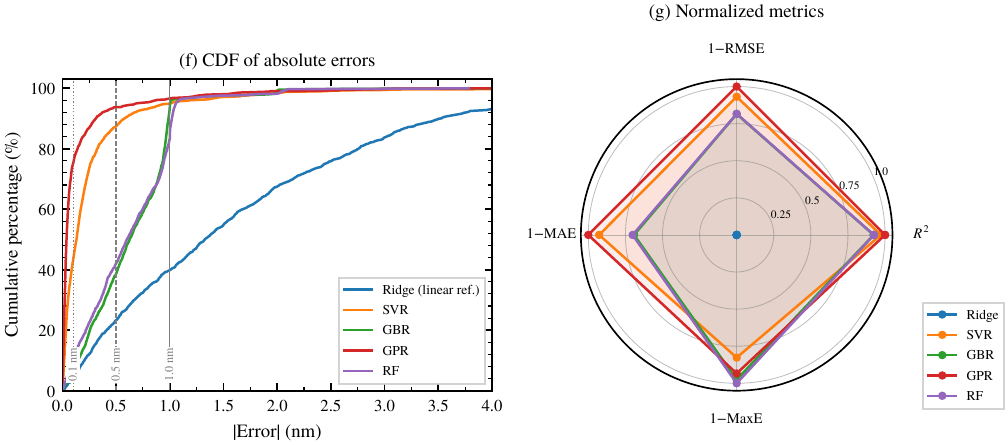}}
    \caption{Performance comparison of five machine-learning models for single-wavelength spectral reconstruction under nested, wavelength-grouped cross-validation (full feature set). (a--e) Estimated versus true wavelength for the out-of-fold estimates of the first repetition (each sample estimated once by a model that never saw its wavelength); dashed line: $y=x$; values in the panels: $R^2$ and RMSE as mean $\pm$ s.d.\ over the 15 outer folds, identical to Table~\ref{tab:Table1}. (f) Cumulative distribution of the absolute errors of all 15 outer folds; reference lines at 0.1, 0.5 and 1~nm. (g) Normalized radar chart of the mean $R^2$, RMSE and MAE over the 15 folds and the maximum error.}
    \label{fig:Fig5}
\end{figure}

The reconstructed versus true wavelengths of the five models are presented in Fig.~\ref{fig:Fig5}(a--e), where the alignment along the diagonal for GPR and SVR contrasts with the systematic deviations of the linear reference at the short-wavelength edge, where the phase changes its slope. To place these results in a unified quantitative context, Fig.~\ref{fig:Fig5}(g) presents a multi-metric radar chart that evaluates all five models on four normalized axes: $1-\mathrm{RMSE}$, $R^2$, $1-\mathrm{MAE}$ and $1-\mathrm{MaxE}$. The enclosed area of each polygon is a visual proxy for the overall model quality. GPR spans the largest area, approaching the outer boundary on all four axes; SVR closely follows, the ensemble methods GBR and RF occupy intermediate positions, and the linear reference collapses toward the origin, particularly on the $1-\mathrm{MaxE}$ and $1-\mathrm{MAE}$ axes, which shows the inadequacy of a linear model for this non-linear spectral mapping. Because the radar chart characterizes average performance, the cumulative error distribution in Fig.~\ref{fig:Fig5}(f) complements it per sample. GPR exhibits the sharpest rise: 90\% of its estimates lie within 0.31~nm of the true wavelength and its median absolute error is 0.04~nm. SVR follows (90\% within 0.58~nm), GBR and RF show broader distributions with tails beyond 1~nm, and the linear reference has a nearly linear distribution across the full error range. Together, the radar chart and the error distribution show that GPR is not only the most accurate model on average but also the most reliable per sample.

\FloatBarrier
\FloatBarrier

The physical origin of this performance becomes clear from a feature ablation, in which seven feature subsets were evaluated with all five models under the same protocol (Supplementary Fig.~S4 and Table~S3). The phase alone outperforms the amplitude alone for every model and in every fold ($p=6\times10^{-5}$), because the phase does not depend on the optical power ($r=-0.005$), whereas the amplitude and the DC voltage scale with it. Adding the amplitude to the phase nearly halves the GPR error (0.67 to 0.37~nm): the spectral shape of the amplitude carries complementary information. The contribution of a single phase feature varies with the modulation frequency, with a single phase feature already giving 0.6--1.2~nm with GPR and the best results at 0.7--1.0~MHz, against about 8~nm for a single amplitude feature (Supplementary Fig.~S3). This frequency dependence is not incidental: different modulation frequencies probe the carrier dynamics on different time scales, so the 15 frequencies together form a multi-dimensional spectral fingerprint, and selecting or weighting the frequencies allows the framework to be optimized for a given accuracy requirement. Measurements at seven power levels make the learned mapping robust to the optical power~\cite{min2024carrier}, which reinforces the advantage of the probabilistic, kernel-based representation of GPR.

Table~\ref{tab:classical} and Supplementary Fig.~S1 compare the learned models with the classical estimators under identical conditions. The analytical group-delay model is limited to 17.8~nm because the measured phase is not proportional to frequency: the amplified detector is a resonant low-pass (Section~\ref{sec:uncertainty}). The single-frequency meter loses accuracy locally wherever $|d\theta/d\lambda|$ at its frequency is small. Using all 15 frequencies removes these blind spots, and the multi-frequency least-squares calibration of the phase reaches 0.65~nm. GPR with the phase alone gives a similar 0.67~nm (Supplementary Table~S3), which validates both approaches; the full-feature GPR reaches 0.38~nm, so machine learning adds value mainly by exploiting the amplitude spectrum jointly with the phase. The linear reference is six times less accurate than GPR, and a second-order polynomial model recovers only part of the gap (1.07~nm), so the inverse problem is non-linear beyond quadratic order, which supports the use of a kernel method.

\FloatBarrier
\begin{table}[!htbp]
\centering
\footnotesize
\begin{tabular}{llcc}
\toprule
Method & Type & RMSE (nm), mean $\pm$ s.d. & MAE (nm)\\
\midrule
\textbf{GPR} & machine learning & $\mathbf{0.383\pm0.148}$ & 0.15\\
SVR & machine learning & $0.518\pm0.217$ & 0.27\\
Multi-frequency least-squares calibration & numerical & $0.654\pm0.301$ & 0.37\\
Random forest & machine learning & $0.743\pm0.126$ & 0.62\\
Gradient boosting & machine learning & $0.746\pm0.127$ & 0.64\\
$k$-NN look-up table & numerical & $0.810\pm0.232$ & 0.64\\
Polynomial Ridge (2nd order) & non-linearity check & $1.07\pm0.38$ & 0.64\\
Ridge (linear ref.) & linear reference & $2.34\pm0.49$ & 1.73\\
Single-frequency calibration & classical OED meter & $7.33\pm6.80$ & 1.79\\
Group-delay model & analytical & $17.8\pm2.7$ & 11.7\\
\bottomrule
\end{tabular}
\caption{Machine-learning models and classical estimators under the identical nested, wavelength-grouped cross-validation (full feature set; the classical calibrations use the phase).}
\label{tab:classical}
\end{table}
\FloatBarrier

Collectively, the evidence of Table~\ref{tab:Table1}, the per-sample scatter plots of Fig.~\ref{fig:Fig5}(a--e), the radar summary of Fig.~\ref{fig:Fig5}(g), the error distribution of Fig.~\ref{fig:Fig5}(f) and the comparison of Table~\ref{tab:classical} lead to a unifying conclusion: the modulated electrical response of a single photodiode encodes sufficient chromatic information to reconstruct the optical wavelength with sub-nanometer accuracy.

\subsection{Dual- and Multi-Line Spectral Reconstruction}
\label{sec:multiline}

We extended the OED-ML framework to the simultaneous reconstruction of two optical wavelengths, $(\lambda_1, \lambda_2)$, from a single photodiode measurement. Two optical tones were combined using a 50:50 polarization-maintaining fiber coupler; one wavelength ($\lambda_1$) was swept across the C band (1530--1565~nm) while the second wavelength ($\lambda_2$) was fixed at 1530, 1540 or 1550~nm. Each sample is characterized by the same 31 features as in the single-wavelength case, with the DC voltage as one scalar, and the target is the wavelength pair. The dataset comprises 213 samples, with the input features in a matrix $\mathbf{X} \in \mathbb{R}^{213 \times 31}$ and the two target wavelengths in a matrix $\mathbf{Y} \in \mathbb{R}^{213 \times 2}$. To estimate the two wavelengths from a common feature set, we employed Scikit-Learn's MultiOutputRegressor, which fits one tuned instance of the base regressor to each wavelength target on the shared features and returns both estimates through a unified interface; target correlations are not modeled explicitly, which gives a clean and consistent framework for simultaneous dual-wavelength estimation~\cite{borchani2015survey,xu2019survey}. The folds are grouped by the swept $\lambda_1$, so every test record is an unseen ordered channel pair (the reversed pair may occur in training), and the hyper-parameters of all five models are selected in the inner loop as for the single-wavelength case (search ranges in Supplementary Section~S1).

For two wavelengths, no single feature type is sufficient (about 3~nm for the phase or amplitude alone; Table~\ref{tab:dual}). The DC voltage alone determines $\lambda_1$ only coarsely and cannot identify $\lambda_2$ (11.0~nm), because the steady-state photovoltage integrates the carrier generation of both superimposed wavelengths into a single scalar. Combined with the phase, however, the DC voltage becomes highly informative: because the power is constant in this experiment, the DC level follows the responsivity at both wavelengths and carries wavelength rather than power information, while the multi-frequency phase supplies the chromatic signature of each tone~\cite{zhang2021dual,darweesh2024nonlinear}. The best set, Phase + DC, gives $0.39\pm0.17$~nm for $\lambda_1$ and $0.58\pm0.23$~nm for $\lambda_2$ with GPR (Table~\ref{tab:dual}) and identifies $\lambda_2$ correctly in every case. The best feature set therefore depends on the task: OED for a single line at variable power, Phase + DC for two lines at fixed power.

\begin{table}[!htbp]
\centering
\footnotesize
\begin{tabular}{lccc}
\toprule
Feature set & $\lambda_1$ (swept) & $\lambda_2$ (three values) & $\lambda_2$ identified\\
\midrule
Phase (15) & 3.38 & 3.78 & 85\%\\
Amplitude (15) & 3.18 & 2.85 & 93\%\\
DC (1) & 2.13 & 11.03 & 34\%\\
OED (30) & 1.04 & 1.22 & 99\%\\
Phase + DC (16) & \textbf{0.39} & \textbf{0.58} & 100\%\\
OED + DC (31) & 0.75 & 0.85 & 99\%\\
\bottomrule
\end{tabular}
\caption{Dual-wavelength reconstruction: GPR RMSE (nm) per feature set, nested cross-validation grouped by $\lambda_1$; mean over the 15 outer folds; last column: the fraction of samples for which the nearest calibrated $\lambda_2$ is the true one. All models: Supplementary Table~S1.}
\label{tab:dual}
\end{table}

\begin{figure}[!htbp]
    \centering
    \adjustbox{max width=\linewidth}{\includegraphics{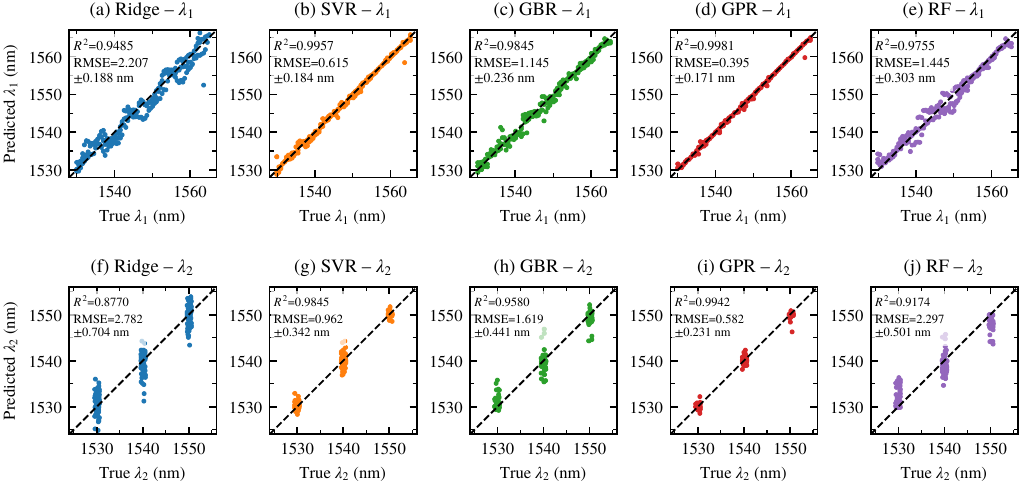}}
    \caption{Dual-wavelength reconstruction: estimated versus true $\lambda_1$ (a--e) and $\lambda_2$ (f--j; horizontal jitter added for visibility) for five models with the best feature set, Phase + DC (16 features; bold row of Table~\ref{tab:dual}), nested cross-validation grouped by $\lambda_1$; points: out-of-fold estimates of the first repetition; values in the panels: $R^2$ and RMSE as mean $\pm$ s.d.\ over the 15 outer folds, identical to Table~\ref{tab:dual} and Supplementary Table~S1. Dashed line: ideal estimate.}
    \label{fig:fig4}
\end{figure}

The reconstruction of all five models with Phase + DC is shown in Fig.~\ref{fig:fig4}, with $\lambda_1$ in panels (a--e) and $\lambda_2$ in panels (f--j); the dashed diagonal represents the ideal estimate, and tighter clustering around it reflects higher fidelity. A clear and consistent ranking emerges for both outputs (mean over the 15 outer folds): GPR is the most accurate (0.39 and 0.58~nm for $\lambda_1$ and $\lambda_2$), followed by SVR (0.62 and 0.96~nm), GBR (1.14 and 1.62~nm), RF (1.45 and 2.30~nm) and the linear reference (2.21 and 2.78~nm). With the full feature set, GPR gives 0.75 and 0.85~nm, and 90\% of its errors lie below 0.63~nm ($\lambda_1$) and 0.81~nm ($\lambda_2$) (Supplementary Table~S1 and Fig.~S2). Across all models the accuracy for $\lambda_2$ is lower than for $\lambda_1$. This asymmetry arises from the different roles of the two components: the swept tone $\lambda_1$ continuously spans the measurement range and produces a well-conditioned mapping, whereas $\lambda_2$ takes only three discrete values and contributes a lower-variance component embedded in the dominant $\lambda_1$ dynamics, so it is less independently observable. The largest GPR errors occur for $\lambda_2=1550$~nm with $\lambda_1=1563.5$~nm (Fig.~\ref{fig:fig4}d,~i). As anticipated for a second wavelength drawn from three values, a $\lambda_2$ value withheld completely requires calibration (5.7--12.8~nm; Supplementary Table~S8); the dual-wavelength result is therefore stated for $\lambda_2$ from a calibrated set, which suits fixed channel plans, and arbitrary spectra are treated next.

\FloatBarrier
\FloatBarrier

In a second campaign, two laser channels were combined before the modulator with both wavelengths swept over 1530--1565~nm, nine separations (1--30~nm) and five power ratios ($-6$ to $+6$~dB). Under nested cross-validation grouped by wavelength pair, GPR reconstructs the spectral centroid of the two lines with 0.36~nm, the longer line with 0.78~nm, the shorter line with 1.37~nm and the separation with 1.51~nm; for random pairs at random powers, never on the calibration grid, these values are 0.70, 1.09, 2.40 and 2.25~nm (Supplementary Figs.~S14 and S15). Lines more than 2~nm apart are resolved, whereas at 1--2~nm the pair responds as a single line. A line with one quarter of the power of the other is still reconstructed, with about twice the error of the stronger line, and the centroid RMSE remains 0.27--0.45~nm over all tested power ratios. A physics-based inversion of the superposition model gives 4.9--11.2~nm for the individual lines on the same data, against 0.8--1.4~nm for GPR, because single-wavelength responses at different wavelengths are very similar in shape and their small differences are captured best by the learned model.

Arbitrary multi-line spectra were studied in a further campaign, in which three laser channels were combined before the modulator. The models were trained on three kinds of data: a single-wavelength library, three-line sweeps with ten spacing patterns and six power patterns, and spectra synthesized from the library by superposition. They were then tested on 200 random spectra that were recorded after the calibration data and never used for training, scaling or model selection. Each test spectrum contains one, two or three lines (29, 72 and 99 spectra) at random wavelengths at least 0.5~nm apart, with random powers of 0.1--0.4~mW per line. The models estimate the line wavelengths; the line powers are not estimated and enter only through the photocurrent weights of the reference centroid (a separate power-ratio test on known wavelength pairs is given in Supplementary Section~S8). In addition to the 31 features, the models receive one physics-based input: the photocurrent-weighted centroid $\bar\lambda=\sum_k P_kR(\lambda_k)\lambda_k/\sum_k P_kR(\lambda_k)$, where $P_k$ is the line power and $R$ the measured responsivity, estimated from the response by a non-negative least-squares fit with the single-wavelength library. This estimate locates the spectrum within the band and leaves the models to resolve the individual lines around it. The centroid itself is reconstructed with sub-nanometer accuracy (0.63~nm on the random spectra and 0.52~nm on unseen three-line combinations of a second session); the detailed centroid results are given in Supplementary Table~S14 and Fig.~S16.

The individual wavelengths of the random spectra are reconstructed with RMSEs of 2.0~nm (two lines) and 3.3~nm (three lines), pooled over all lines, against 8.7 and 6.9~nm for a reference that ignores the measurement and always returns the average line positions (Supplementary Fig.~S17). A deep neural network trained on both sessions with synthetic augmentation gives 1.77--2.37~nm per line (2.03~nm overall) on three-line combs withheld in grouped cross-validation and 3.31~nm on the 99 random three-line spectra (Supplementary Table~S15). When the two line spacings are known, as for a fixed channel grid such as a WDM comb, only the position of the comb is estimated, and all three lines are reconstructed with 0.70~nm.

\begin{figure}[!htbp]
    \centering
    \adjustbox{max width=\linewidth}{\includegraphics{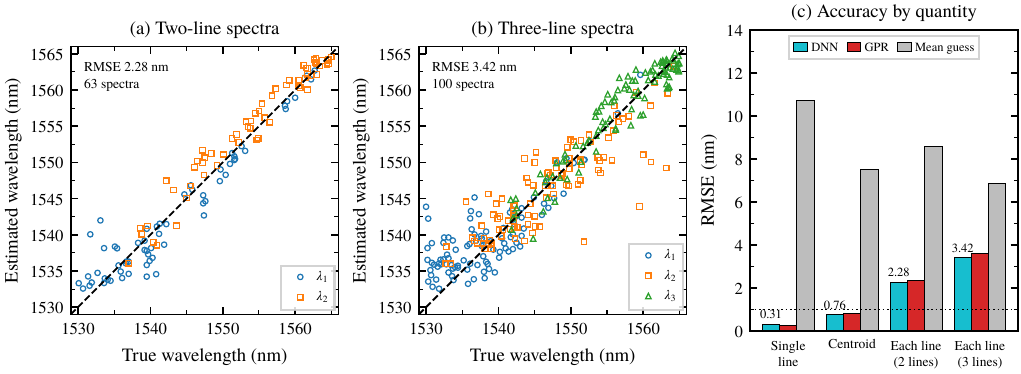}}
    \caption{Multi-line spectral reconstruction with one photodiode in the independent campaign (three laser channels combined before the modulator; 200 random spectra of one session, models trained on another session): estimated versus true wavelength of each line, ordered $\lambda_1<\lambda_2<\lambda_3$, in (a) two-line and (b) three-line spectra (DNN); (c) RMSE of the single line, the centroid and the individual lines for DNN and GPR, compared with a reference that always returns the average line positions (mean guess); dotted line: 1~nm.}
    \label{fig:multiline}
\end{figure}

The multi-line results were confirmed in an independent campaign recorded in three sessions, in which a reference spectrum was measured periodically and every acquisition was referred to the last preceding reference of the same session (Section~\ref{sec:setup}). Only acquisitions recorded in one stable instrument configuration were used (Supplementary Section~S7). The features are the 15 phases and 14 logarithmic amplitudes of the response normalized to its 0.1-MHz amplitude, which removes the optical power. Models were trained in one session on the single-line library, three-line sweeps and spectra synthesized from that library, and tested on 200 random spectra (one to three lines, random wavelengths and powers) of another session, with the number of lines given. The DNN reconstructs each line of two-line and three-line spectra with 2.28 and 3.42~nm (GPR: 2.33 and 3.62~nm), against 8.6 and 6.9~nm for the reference that always returns the average line positions, and a single line with 0.31~nm (GPR: 0.26~nm) (Fig.~\ref{fig:multiline}).

\FloatBarrier
\FloatBarrier

The accuracy for individual lines follows from the information content of one measurement. The multi-line response is the power-weighted sum of single-wavelength responses, so every measurement lies in the space spanned by the single-wavelength library. A singular-value decomposition of the library (351 wavelengths on a 0.1-nm grid; each of the 31 features divided by its measured noise) gives the number of independent directions of this space above the noise (Supplementary Fig.~S18): the RMS signal along the first five directions is 72, 14.8, 2.5, 0.71 and 0.25 times the noise of one measurement. Three directions clearly exceed the noise and a fourth is comparable with it, so one measurement determines about three to four numbers, whereas a spectrum of $N$ lines of unknown power has $2N$ unknowns. The wavelength of one line and the centroid of several lines are carried by the strongest directions and are reconstructed with sub-nanometer accuracy; the individual lines of two- and three-line spectra draw on the weaker directions and are reconstructed at the nanometer scale. An exhaustive maximum-likelihood search over all three-line combinations of the measured library gives 7.1~nm for the individual lines of the 99 random three-line spectra, so the learned models (2.0--3.3~nm) extract considerably more than direct library matching; a simulation with the measured noise indicates that a 100-fold lower noise would bring the individual lines to about 1~nm. This information content is a property of the present configuration, not a fundamental limit of OED: lower noise, more modulation frequencies or additional independent channels, such as OED measured at several bias voltages~\cite{dutta2024large} or detectors with different absorption edges~\cite{yoon2022miniaturized}, add independent directions and are expected to improve the individual-line accuracy further. The present demonstration concerns spectra composed of discrete lines; broadband (continuous) spectra have many more unknowns and call for dedicated experiments.

Because the lines are combined before one modulator, they share the fiber mode, the illuminated detector area and the modulation, so overlap and synchronization are fixed by design. Lines at least 0.5~nm apart beat at $\geq$62~GHz, far above the detector bandwidth, so their photocurrents add incoherently. A dedicated test with three wavelength pairs, five power ratios from 1:4 to 4:1 and eight repeats shows that the dual response is approximately the sum of the single-wavelength responses (RMS residual of the eight-repeat mean: 0.32$^\circ$ in phase and 0.95\% in amplitude), and each line is weighted by its power (fitted weight within 0.026 of the set power fraction; Supplementary Fig.~S19). Equal intensities are therefore not required; a weaker line, or two closely spaced lines, leaves a smaller signature. With separate modulators, synchronization would matter: at 1.5~MHz a differential delay of 1.9~ns shifts the relative modulation phase of the two lines by about 1$^\circ$, equivalent to at most about 1.6~nm; the single-modulator design avoids this.

\subsection{Residual Analysis and Error Distribution}

\begin{figure}[!htbp]
    \centering
    \begin{subfigure}{\linewidth}
        \centering
        \adjustbox{max width=\linewidth}{\includegraphics{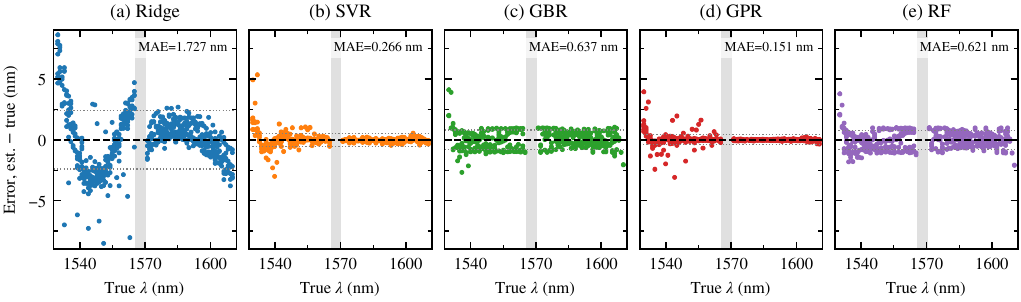}}
       \caption{}
    \end{subfigure}
    \vspace{0mm}
    \begin{subfigure}{\linewidth}
        \centering
        \adjustbox{max width=\linewidth}{\includegraphics{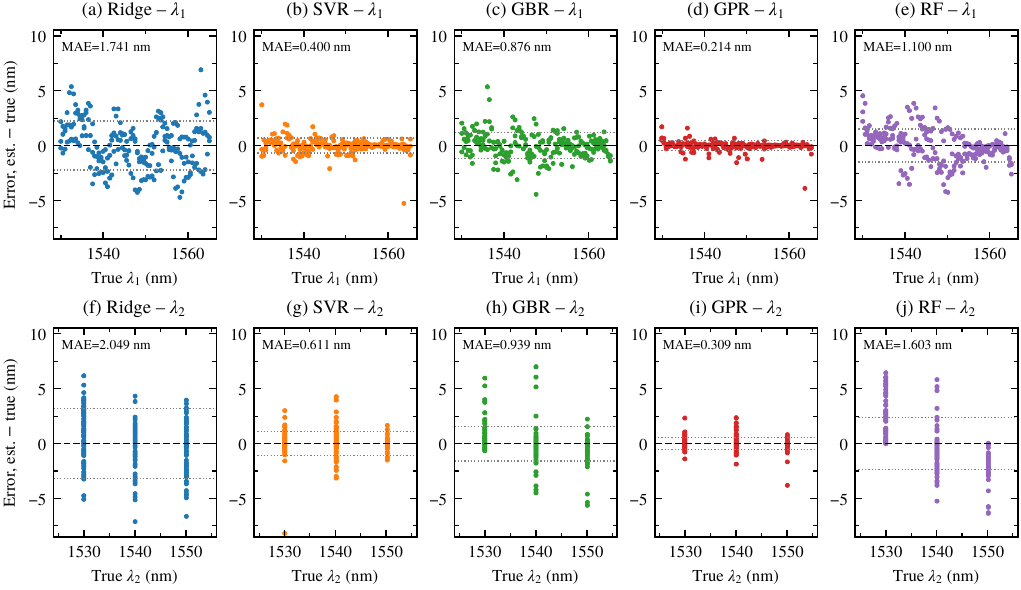}}
        \caption{}
    \end{subfigure}
    \caption{Residual analysis of spectral reconstruction under nested, wavelength-grouped cross-validation (points: out-of-fold estimates of the first repetition; MAE in the panels: mean over the 15 outer folds, as in the tables; full feature set in (a), Phase + DC in (b), as in Figs.~\ref{fig:Fig5} and \ref{fig:fig4}). (a) Single-wavelength residuals (estimated $-$ true) versus true wavelength; dashed: zero, dotted: $\pm1$ s.d.\ of the plotted errors; grey: not measured. (b) Dual-wavelength residuals of $\lambda_1$ (top) and $\lambda_2$ (bottom).}
    \label{fig:combined_residuals}
\end{figure}

We evaluated the residuals $\Delta\lambda=\lambda_\mathrm{est}-\lambda_\mathrm{true}$ (estimated minus true, the sign convention used in all figures) across the 1530--1610~nm band (Fig.~\ref{fig:combined_residuals}). As shown in Fig.~\ref{fig:combined_residuals}(a), all regressors capture the global trend of the OED, but their precision varies with the local non-linearity of the photodiode response. The linear reference has the largest spread (MAE 1.73~nm), failing to follow the non-linear phase--wavelength relationship; SVR reduces the spread (MAE 0.27~nm) and RF and GBR provide robust non-linear fitting (MAE 0.62 and 0.64~nm), while GPR achieves the highest fidelity (MAE 0.15~nm) by modeling the smooth underlying structure of the OED response. For two-wavelength spectra (Fig.~\ref{fig:combined_residuals}b, Phase + DC), GPR and SVR remain centred on zero for both outputs (MAE 0.21 and 0.31~nm for GPR), whereas the linear reference, RF and GBR show larger scatter.

\FloatBarrier
\FloatBarrier

\begin{figure}[!htbp]
    \centering
    \adjustbox{max width=\linewidth}{\includegraphics{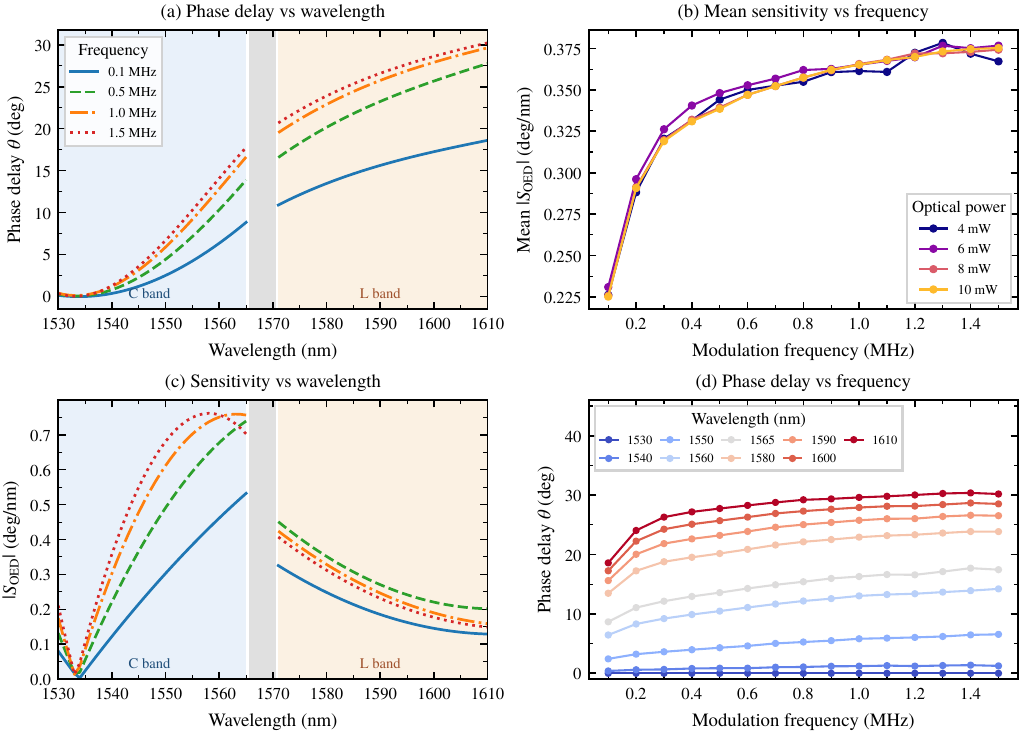}}
    \caption{OED characterization of the Ge photodiode across the C+L band (single-wavelength dataset; curves in (a)--(c) smoothed with a cubic smoothing spline in each band). (a) Wavelength-dependent phase delay at four modulation frequencies (mean over the power levels). (b) Mean OED sensitivity over 1530--1610~nm versus modulation frequency for four optical powers, demonstrating power-independent behaviour. (c) OED sensitivity $|S_\mathrm{OED}|$ versus wavelength at four modulation frequencies. (d) Phase delay versus modulation frequency for wavelengths from 1530 to 1610~nm. Blue: C band, orange: L band, grey: not measured.}
    \label{fig:fg8}
\end{figure}

The primary physical mechanism governing the spectral distribution of the residuals is the wavelength-dependent information of the OED, quantified by the phase sensitivity $d\theta/d\lambda$ (Fig.~\ref{fig:fg8}c). A larger $|d\theta/d\lambda|$ means that adjacent wavelengths produce a greater differential phase response across the 15 modulation channels, generating a more separable feature representation from which any regression model can learn a more accurate inverse mapping. Because the electronic noise floor, dominated by thermal Johnson--Nyquist noise and photocurrent shot noise, is spectrally flat and wavelength-independent~\cite{agrawal2012fiber,quinlan2013analysis}, the wavelength-discrimination signal-to-noise ratio scales as $\mathrm{SNR}(\lambda)\propto|d\theta/d\lambda|$. According to the OED framework of~\cite{glasser2021optoelectronic}, this sensitivity is proportional to the spectral absorption term $\alpha^{-1}d\alpha/d\lambda$, which encodes the wavelength dependence of carrier diffusion and transit within the photodiode. The measured response is summarized in Fig.~\ref{fig:fg8}: the phase delay at four modulation frequencies (a), the mean sensitivity at four power levels (b), the sensitivity $|S_\mathrm{OED}|=|d\theta/d\lambda|$ versus wavelength (c) and the phase versus frequency (d).

The phase increases monotonically with wavelength but not linearly: slowly near 1530~nm, steeply at 1545--1565~nm and more gradually in the L band (Fig.~\ref{fig:fg8}a). The sensitivity of the smoothed curves reaches 0.76$^\circ$/nm near 1558~nm at 1.5~MHz (Fig.~\ref{fig:fg8}c; 0.85$^\circ$/nm at 1561~nm before smoothing). Unsmoothed, the median of its maximum over frequency is 0.73$^\circ$/nm at 1535--1565~nm, 0.29$^\circ$/nm at 1571--1610~nm and 0.13$^\circ$/nm at 1530--1534~nm. In the short-wavelength region the sensitivity is low and changes rapidly, so small phase errors translate into larger wavelength errors, and the residuals of all models, particularly of the linear reference, are largest there (Fig.~\ref{fig:combined_residuals}a). This behaviour is consistent with the independent measurements of Liokumovitch et al.~\cite{liokumovitch2021optoelectronic}, who reported a peak OED sensitivity of about 1$^\circ$/nm at 4~MHz in the same spectral window. In the L band the residuals are smallest; in the C band, where the sensitivity is highest, they are somewhat larger, which is attributed to slow drift of the steep absorption edge (Section~\ref{sec:uncertainty}).

The sensitivity curves of the different power levels coincide (Fig.~\ref{fig:fg8}b), confirming that the encoding is governed by the wavelength-dependent absorption depth rather than by the optical power~\cite{lamminpaa2006characterization}.

The phase grows with modulation frequency for every wavelength (Fig.~\ref{fig:fg8}d): higher modulation frequencies produce larger phase separations between adjacent wavelengths, amplifying the discriminative information of the measurement, which is reflected in the monotonic increase of the mean sensitivity from 0.23$^\circ$/nm at 0.1~MHz to 0.37$^\circ$/nm at 1.5~MHz (Fig.~\ref{fig:fg8}b). Because the detector signal decreases above its bandwidth, this trade-off sets the useful band (Section~\ref{sec:uncertainty}).

The OED is largest where the absorption depth $1/\alpha$ changes fastest. As the wavelength increases, the Ge absorption coefficient $\alpha$ decreases and photons penetrate progressively deeper into the device~\cite{lamminpaa2006characterization}, which steepens $\theta(\lambda)$. In Ge at 300~K, $1/\alpha$ grows 25-fold, from 2.7 to 67~$\mu$m, between the direct-gap threshold (0.81~eV, about 1531~nm~\cite{dash1955intrinsic}) and 1610~nm~\cite{nunley2016optical}. The measured sensitivity follows this spectrum: small below about 1535~nm, where the absorption is strong and nearly constant, largest near 1555--1565~nm, where $\alpha$ falls fastest, and decreasing in the L band as the spectral gradient $|\alpha^{-1}d\alpha/d\lambda|$ declines (Supplementary Fig.~S22). The measured responsivity of our photodiode decreases smoothly from 0.96~A/W at 1530~nm to 0.52~A/W at 1608~nm (Supplementary Fig.~S23). In summary, the residuals across the C+L band are governed by the interplay between the wavelength-dependent OED sensitivity and a spectrally flat electronic noise floor, together with the stability of the absorption edge, so the performance ceiling of any machine-learning model in OED-based spectroscopy is set by the discriminative information encoded in the carrier-transport physics of the device rather than by the regression architecture~\cite{glasser2021optoelectronic,liokumovitch2021optoelectronic}.

\FloatBarrier
\FloatBarrier

\subsection{Nested K-Fold Cross-Validation for High-Precision Wavelength Reconstruction}
\label{sec:cv}

To evaluate how well each model performs on unseen wavelengths, nested, wavelength-grouped five-fold cross-validation repeated three times (Section~\ref{sec:sec_3.1}) was conducted for the five regression models. The per-fold RMSE and MAE are presented in Fig.~\ref{fig:cv_overview}, and the fold-resolved heatmap in Supplementary Fig.~S5. The results show a clear and consistent ranking. GPR delivers the best performance by a notable margin, with an RMSE of $0.383\pm0.148$~nm and an MAE of $0.151\pm0.060$~nm; it has the lowest RMSE in 13 of the 15 outer folds (SVR in folds 4 and 6) and stays below 0.7~nm in every fold. SVR is second ($0.518\pm0.217$~nm), with a larger fold-to-fold spread, followed by RF ($0.743\pm0.126$~nm) and GBR ($0.746\pm0.127$~nm), which behave steadily across folds, and the linear reference ($2.34\pm0.49$~nm), whose errors persist in every fold as a direct result of its linear structure. This outcome is grounded in the physics of the problem: the phase response $\theta(\lambda)$ carries wavelength information through smooth but complex non-linear variations, and the non-parametric Bayesian formulation of GPR~\cite{rasmussen2006gaussian} learns this mapping directly from the observations without a fixed functional form. GPR is also the slowest model to train (2.8~s for 532 samples), with 14~$\mu$s per estimate (Supplementary Fig.~S6).

\FloatBarrier
\begin{figure}[!htbp]
    \centering
    \begin{subfigure}[b]{0.95\linewidth}
        \centering
        \adjustbox{max width=\linewidth}{\includegraphics{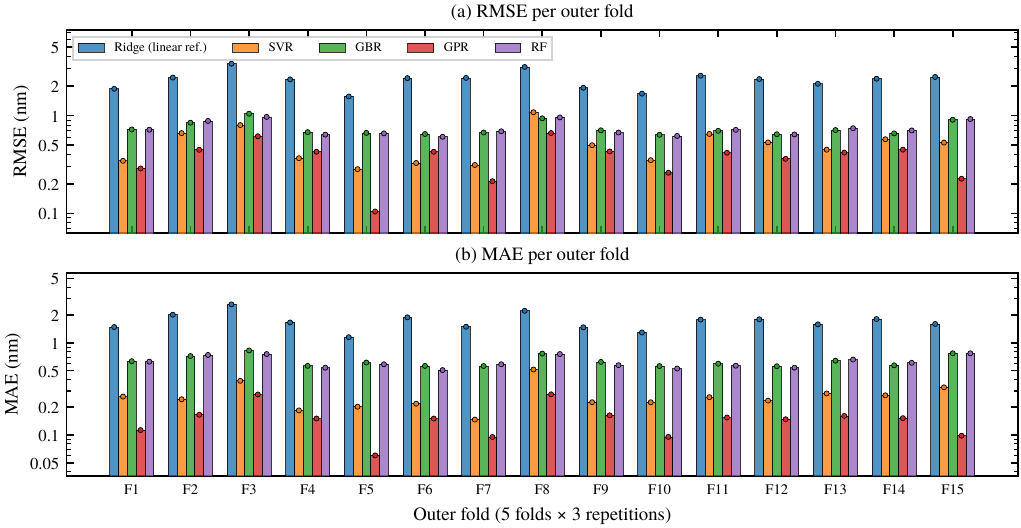}}
        \caption{}
        \label{fig:cv_per_fold}
    \end{subfigure}
    \vspace{0.5em}
    \begin{subfigure}[b]{0.9\linewidth}
        \centering
        \adjustbox{max width=\linewidth}{\includegraphics{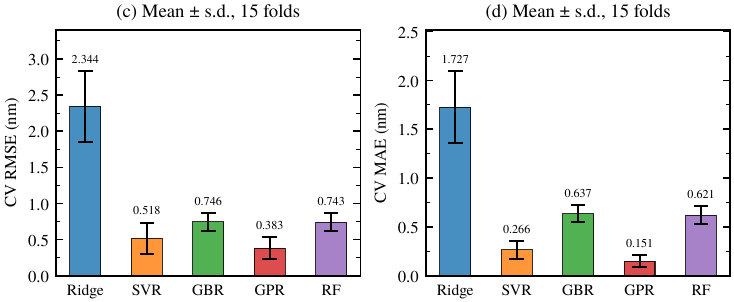}}
        \caption{}
        \label{fig:cv_mean_std}
    \end{subfigure}
    \caption{Nested, wavelength-grouped cross-validation of five regression models (5 folds $\times$ 3 repetitions). (a) RMSE and MAE in each of the 15 outer folds. (b) Mean RMSE (left) and MAE (right) with one standard deviation across folds: GPR reaches $0.383\pm0.148$~nm (MAE $0.151\pm0.060$~nm), and the linear reference $2.34\pm0.49$~nm. The fold-resolved heatmap is shown in Supplementary Fig.~S5.}
    \label{fig:cv_overview}
\end{figure}
\FloatBarrier

To test more than interpolation between 1-nm neighbours, windows of 1--13 consecutive wavelengths were withheld; for each width the windows tile the band without overlap, at least 2~nm from the band edges and the unmeasured gap, and the RMSE is pooled over all withheld samples (Supplementary Fig.~S8b and Table~S5). GPR stays at 0.35--0.72~nm, whereas the least-squares calibration degrades from 0.56 to 3.72~nm, because a spline per frequency cannot bridge a wide gap while GPR learns the joint mapping from all features. When contiguous 10-nm blocks are withheld (Supplementary Fig.~S8a and Table~S4), the error inside the band is 0.05--1.09~nm; for the two edge blocks, where the model extrapolates, it is 2.8 and 1.5~nm, so, as for any calibrated instrument, the calibration should cover the operating band.

When one complete power level is withheld together with grouped cross-validation over the wavelengths (Supplementary Fig.~S8c and Table~S6), the interior levels 5--9~mW give 0.22--0.50~nm; the upper edge level (10~mW) gives 0.35~nm and the lower edge level (4~mW) 2.5~nm, so the calibration should span the power range of use.

The models trained on the complete dataset were applied without change to 60 acquisitions of a separate session at three previously calibrated wavelengths (1540, 1550 and 1560~nm), with that session's own 1530-nm phase reference (Supplementary Fig.~S9 and Table~S7). The phase-only GPR transfers directly with 0.60~nm and a mean offset of $+0.51$~nm; the repeats at one wavelength scatter by only 0.02--0.10~nm, so the error is mainly a session-dependent offset. The OED set gives 0.92~nm and the OED + DC set 12.89~nm, because the DC level of that session was 0.42--0.45 times that of the calibration. Within one session the OED and OED + DC sets are the most accurate; across sessions the phase features transfer directly, and the amplitude and DC can be included after a short reference measurement in the new session, as in the independent multi-line campaign (Section~\ref{sec:multiline}).

\subsection{Spectral Error, Uncertainty and Resolution Analysis for the GPR Model}
\label{sec:uncertainty}

\begin{figure}[!htbp]
    \centering
    \adjustbox{max width=\linewidth}{\includegraphics{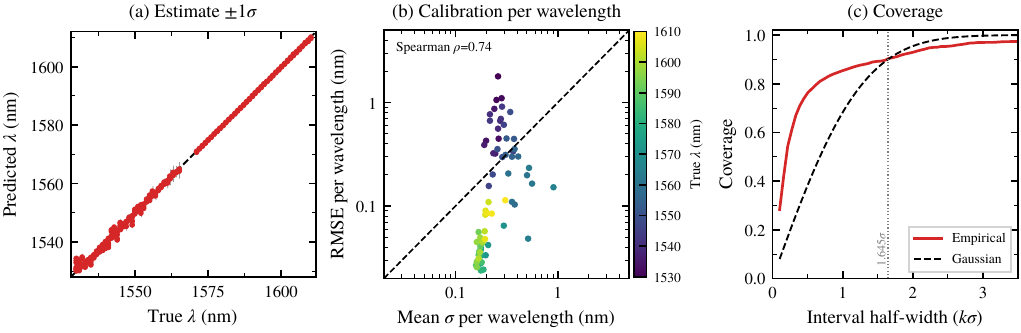}}
    \caption{Uncertainty quantification of the Gaussian process regression (GPR) model for wavelength reconstruction under nested, wavelength-grouped cross-validation (all 15 outer folds). (\textbf{a}) Estimated versus true wavelength with $\pm1\sigma$ error bars; the dashed line represents perfect agreement. (\textbf{b}) RMSE per wavelength versus the mean predicted $\sigma$, colour-coded by wavelength; dashed line: equality; Spearman $\rho=0.74$. (\textbf{c}) Empirical coverage of the $\pm k\sigma$ interval compared with the theoretical Gaussian expectation as a function of the interval half-width.}
    \label{fig:gpr_uncertainty}
\end{figure}

To assess the accuracy and reliability of the uncertainty estimates, we performed uncertainty quantification of the GPR model, which returns a predictive standard deviation $\sigma$ with every estimate (Fig.~\ref{fig:gpr_uncertainty}). The estimates follow the ideal line across the examined spectral range (1530--1610~nm) without systematic bias, with narrow $\pm1\sigma$ intervals (panel a). The calibration of the uncertainty is validated in panel (b), which compares the error per wavelength with the predicted uncertainty: a statistically significant positive Spearman correlation ($\rho=0.74$, $p=3\times10^{-14}$) confirms that the predicted uncertainty reflects the actual error magnitude, with larger $\sigma$ and error in the C band, where the sensitivity changes fastest, and at the short-wavelength edge. The empirical coverage of the $1.645\sigma$ interval is 90\% (nominal 90\%; panel c). At narrow intervals the coverage exceeds the Gaussian expectation (85\% within $1\sigma$, nominal 68\%), so the model is slightly conservative, and wider intervals slightly under-cover (93\% within $2\sigma$) because of a few larger errors at the short-wavelength edge. The predicted standard deviation refers to a noisy observation (it includes the learned noise term) and indicates where an estimate can be trusted, which is valuable for uncertainty-aware inverse design and real-time spectral reconstruction in integrated photonic applications. The trade-off between computational efficiency and accuracy of the five models is analyzed in Supplementary Section~S4.

\FloatBarrier
\FloatBarrier

The resolution for a single line is set by the phase noise and the OED sensitivity. A wavelength change $\delta\lambda$ shifts the phase by $(d\theta/d\lambda)\,\delta\lambda$, so the resolution at one frequency is $\delta\lambda\approx\sigma_\theta/|d\theta/d\lambda|$; assuming independent phase noise at the 15 frequencies, the frequencies combine in an approximate local Cram\'er--Rao bound (CRB) for an unbiased estimator that uses the phase only. The phase noise estimated from the scatter between acquisitions is 0.06--0.08$^\circ$ at 0.1--0.7~MHz and rises to 0.19$^\circ$ at 1.5~MHz as the detector response falls. With the measured sensitivities, this approximate bound has a median of 0.071~nm, 2.2 times lower than for the best single frequency (Table~\ref{tab:crb}).

For comparison with this bound, the error per wavelength is the RMSE over all 21 estimates of that wavelength (seven powers $\times$ three repetitions). The GPR error follows the approximate bound (median ratio 1.0) and is lower than the least-squares error at 61 of the 76 wavelengths, whereas the least-squares calibration stays about three times above it (Supplementary Fig.~S13). In the L band GPR reaches the bound and at some wavelengths falls below it; this is possible because the bound uses only the phase and assumes independent frequencies and unbiased estimates, whereas GPR also uses the amplitude and DC, so the curve is a reference rather than a strict limit for GPR. In the C band the error is about four times the bound, which points to drift as the main remaining contribution.

\begin{table}[!htbp]
\centering
\footnotesize
\begin{tabular}{lccc}
\toprule
 & 1530--1534~nm & 1535--1565~nm & 1571--1610~nm\\
\midrule
Sensitivity (median of the maximum over $f$) & 0.13$^\circ$/nm & 0.73$^\circ$/nm & 0.29$^\circ$/nm\\
Bound, best single frequency & 1.67~nm & 0.21~nm & 0.14~nm\\
Bound, 15 frequencies combined & 0.70~nm & 0.074~nm & 0.065~nm\\
Achieved GPR error (median per wavelength) & 1.06~nm & 0.32~nm & 0.037~nm\\
\bottomrule
\end{tabular}
\caption{Wavelength sensitivity (unsmoothed derivatives), approximate phase-only resolution bound and achieved error in three spectral regions.}
\label{tab:crb}
\end{table}

A plausible contributor to the difference between the C and L bands is slow drift of the detector response, measured as an equivalent wavelength drift of 0.08--0.14~nm/h at 1550~nm. The steep absorption edge that gives the C band its high sensitivity also makes it sensitive to temperature: the Ge band gap shifts by about $-4\times10^{-4}$~eV/K, i.e.\ about 0.8~nm/K at 1550~nm, so a temperature change of a fraction of a kelvin per hour could explain the drift; its contribution was not measured directly. The laser drifts by less than 4~pm/h. Table~\ref{tab:budget} compares all uncertainties with the reconstruction error: the wavelength reference and the measurement repeatability (Supplementary Figs.~S11 and S12) are 6--700 times smaller than the error, so the reported accuracy reflects the method itself, and the remaining error is attributed to drift and to the calibration model between the 1-nm calibration points; periodic reference measurements, as used in the independent multi-line campaign, correct the drift.

\begin{table}[!htbp]
\centering
\footnotesize
\begin{tabular}{p{8.4cm}cc}
\toprule
Source & Size & Reconstruction error / source\\
\midrule
Laser set-point repeatability (optical spectrum analyzer) & 0.5--0.6~pm & $\sim$700\\
Laser repeatability (datasheet) & $\pm$2.5~pm & 150\\
Laser absolute accuracy (datasheet; common to calibration and test) & $\pm$22~pm & 17\\
OED repeatability between sessions & 12--61~pm & 6--30\\
Approximate phase-only Cram\'er--Rao bound (median over the band) & 71~pm & 5\\
Slow drift of the detector response & 0.08--0.14~nm/h & ---\\
\midrule
\textbf{Reconstruction error (GPR, nested wavelength-grouped CV)} & \textbf{0.38~nm} & \\
\bottomrule
\end{tabular}
\caption{Error budget of the single-wavelength reconstruction.}
\label{tab:budget}
\end{table}

Germanium has no narrow absorption resonances in the 1530--1610~nm band at room temperature: the absorption is governed by the direct-gap edge and, at longer wavelengths, by the weaker indirect gap, and excitonic features are thermally broadened. The literature absorption spectrum of bulk Ge and the measured responsivity of our device are smooth (Supplementary Figs.~S22 and S23); neither was measured in the 1566--1570~nm gap, and only the trends on both sides are compared. Any reproducible structure, such as the edge itself or a weak ripple from optical components, is learned by the calibration and reproduces between sessions (Supplementary Fig.~S24); slow non-reproducible change, chiefly the temperature shift of the absorption edge, is addressed by temperature stabilization of the photodiode, periodic referencing with a known wavelength and recalibration after a change of the set-up. RF interferometric phase-shift amplification has reached femtometer resolution with the same effect in a single-wavelength monitor~\cite{liokumovitch2022femtometer}.

The spectral range also follows from the absorption: in the present Ge photodiode with optical-intensity modulation, the sensitivity follows the change of the absorption depth with wavelength. Below the direct-gap threshold (about 1531~nm) the absorption is strong and nearly constant, so the sensitivity falls to 0.13$^\circ$/nm and the bound rises to 0.7~nm, which is why the error is largest at the short end of the band. The range of this demonstration (1530--1565 and 1571--1610~nm, with 1610~nm the highest measured setpoint) is set by the two laser modules; wavelengths above 1610~nm were outside the measurement range. These values describe this Ge photodiode with intensity modulation; other detectors or modulation schemes can extend the operating range, and the operating range of a calibrated instrument equals its calibrated range.

The modulation band is set by the detector. Its response, measured with the VNA at 1535, 1550 and 1565~nm in two independent runs, is that of a resonant second-order low-pass with $f_0\approx333$~kHz and $Q\approx1.5$; at 1.5~MHz it is 4.3\% of the low-frequency value, and the sweep-to-sweep phase noise rises from about 0.005$^\circ$ at 0.1~MHz to 0.06--0.09$^\circ$ at 1.5~MHz, while the mean sensitivity at 1535--1565~nm rises from 0.29 to 0.62$^\circ$/nm (Supplementary Fig.~S20). The band of 0.1--1.5~MHz therefore covers the full rise of the OED sensitivity while the signal is still measurable: the upper limit is the detector bandwidth, beyond which the gain in sensitivity no longer compensates the loss of signal, and the lower limit is that the OED phase vanishes as $f\rightarrow0$. These limits belong to this detector rather than to the method: a faster amplifier would extend the band to higher frequencies, where the OED is larger; a large OED has been measured in Ge photodiodes at 10~MHz~\cite{dutta2024large}.

With respect to the modulation waveform, the detector is linear in the operating range: the log--log slope of the RF amplitude versus optical power is 1.01 (median; interquartile range 0.99--1.02), and the phase does not depend on power (Supplementary Fig.~S21). For a linear system any periodic modulation is a sum of sinusoids at the fundamental and its harmonics, each experiencing the same transfer function, so no particular waveform is required; the waveform only determines which frequencies one measurement probes and how much signal each receives. The resolution of one measurement predicted from this linear model and the measured VNA noise is 11--19~pm for sine, square, ramp and pulse modulation (Supplementary Table~S18). Non-sinusoidal waveforms measure many frequencies at once and require harmonic detection; we used a swept sine, which narrowband VNA detection measures with the lowest noise at each frequency.

Germanium is cubic and optically isotropic, so the absorption depth, and with it the OED, should not depend on polarization; residual polarization-dependent loss changes only the power, on which the phase does not depend. With a polarization controller between the modulator and the photodiode (1550~nm, 20~mW laser setting, 0.5~MHz), 40 polarization states were measured in four independent runs (Supplementary Fig.~S25). The spread of the state means (0.15$^\circ$) does not exceed the spread expected from reading noise alone (0.17$^\circ$), the states do not differ significantly (one-way ANOVA, $p=0.70$), and the amplitude changes by less than 1\%. The 95\% upper limit of a polarization-induced error is 0.20~nm, about half the reconstruction error, so in this configuration standard non-polarization-maintaining fiber can be used in front of the detector.

\section{Conclusion}

In summary, we demonstrate discrete multi-line spectral reconstruction with a single germanium PN photodiode that uses its intrinsic optoelectronic chromatic dispersion (OED) as the spectral encoder, without dispersive optics, filter arrays or detector arrays. To our knowledge, this is the first demonstration of multi-line spectral reconstruction with a single photodiode using OED information. The DC photovoltage and the RF amplitude and phase at 15 modulation frequencies are mapped to the spectrum by machine learning and evaluated under a nested, wavelength-grouped validation that excludes spectral leakage. Across the C and L bands (1530--1610~nm), Gaussian process regression reconstructs the wavelength of a single line with an RMSE of 0.38~nm, compared with 0.65~nm for the best classical calibration, and remains accurate for withheld wavelength ranges and in a separate session. The feature ablation establishes phase-dominant spectral encoding: the RF phase carries most of the wavelength information because it does not depend on optical power, and the amplitude spectrum adds complementary information. For two lines, the swept wavelength is reconstructed with 0.39~nm when the second line belongs to a calibrated set. For mixtures with randomly varied wavelengths and powers, the pooled RMSE of the estimated line positions is 2.0~nm for two-line spectra and 3.3~nm for three-line spectra, two to four times lower than a reference that ignores the measurement, and an independent campaign confirms these results across sessions. Broadband (continuous) spectra are a natural next step. The singular-value analysis links the individual-line accuracy to the number of independent response directions above the noise, so additional measurement channels, such as OED at several bias voltages or detectors with different absorption edges, together with lower noise, offer a clear path to higher spectral resolution. The single-detector architecture therefore provides a promising route toward compact, alignment-free and potentially monolithic spectral sensors for optical communications, environmental monitoring and portable spectroscopy.

\section*{Author contributions}

\noindent
\textbf{Endalamaw Ewnu Kassa:} Conceptualization, Methodology, Software, Validation, Formal analysis, Investigation, Data curation, Visualization, Writing -- original draft, Writing -- review \& editing.

\noindent
\textbf{Ziv Glasser:} Conceptualization, Methodology, Investigation, Resources, Validation, Writing -- review \& editing.

\noindent
\textbf{Uttama K. Saint:} Methodology, Validation, Formal analysis, Writing -- review \& editing.

\noindent
\textbf{Roi Yozevitch:} Conceptualization, Methodology, Software, Supervision, Writing -- review \& editing.

\noindent
\textbf{Shmuel Sternklar:} Conceptualization, Methodology, Investigation, Validation, Resources, Supervision, Project administration, Writing -- review \& editing.

\section*{Competing interests}

The authors declare that they have no known competing financial interests or personal relationships that could have appeared to influence the work reported in this paper.

\section*{Acknowledgements}

The authors acknowledge the Electro-Optics Laboratory at Ariel
University for providing the laboratory facilities, equipment, and technical
resources used in this research.

\section*{Funding}

This research was supported by the Israel Science Foundation (Grant No. 1215/26). The research was conducted using facilities and resources provided by the Electro-Optics Laboratory at Ariel University.

\section*{Data availability}

Data will be made available on request.

\bibliographystyle{unsrtnat}
\bibliography{references}

\end{document}